\documentclass{ws-ijmpa}
\usepackage[compress]{cite}
\usepackage{graphicx}
\usepackage{graphics}
\usepackage{bookmark}

\begin{document}
\title{NEUTRINO-ANTINEUTRINO ASYMMETRY FROM THE SPACE-TIME NONCOMMUTATIVITY}
\author{Ouahiba MEBARKI$^{1,2}$, N. MEBARKI$^{2}$ and H. AISSAOUI$^{2}$}
\address{1.Department of Physics, University of 20 Ao\^ut 1954,  Skikda, Algeria \footnote{Permanent address}\\
2. Laboratoire de Physique Math\'{e}matique et Subatomique (LPMS),\\ 
Constantine 1 University, Constantine, Algeria \\ mebarki\_ouahiba@yahoo.fr}

\maketitle
\begin{abstract}
A new mechanism having as an origin the space-time non commutativity has
been shown to generate anisotropy and axial like interaction giving rise to
a leptonic asymmetry for fermionic particles propagating in a curved non
commutative $FRW$ universe. As a by product, for ultra relativistic
particles like neutrinos, an analytical expression of this asymmetry is
derived explicitly. Constraints and bounds from the cosmological parameters
are also discussed.
\end{abstract}

\keywords{Noncommutative Geometry; Modified Theories of Gravity; Leptogenesis.}

\section{Introduction}
\label{sec1}

 Leptogenesis is a very important observation and can affect the
present energy density and cosmic microwave background ($CMB$) etc...\cite{1}%
. To explain the origin of this lepton number asymmetry, many mechanisms
were proposed like the one of Afflek-Dine \cite{2}, fermion propagation in a
curved space-time \cite{3,4}, Lorentz and $CPT$ violating scenarios in the
context of Riemanian-Cartan space-time \cite{5,6} etc... On the other hand,
the non commutative nature of the space-time has been a subject of a very
active research \cite{7,8,9,10,11,12,13,14,15,16,17}. The motivation was
that the space-time non commutativity could be significant in the early
universe where quantum gravity effects become important and very sensitive
to search for signatures in the cosmological observations \cite{18}.
Moreover, the observed anisotropies of the $CMB$ may be caused by the non
commutativity of space-time geometry \cite{19,20}. The main goal of this
paper is to show that a dynamical leptonic (including neutrinos) asymmetry
can be generated only by the non commutativity of a curved expanding
space-time leading to a new mechanism explaining matter-antimatter asymmetry
in the universe. In section \ref{sect2}, we present the mathematical
formalism. In section \ref{sect3},we derive the analytical expression of the
neutrino-antineutrino asymmetry. Finally, in section \ref{sect4}, we discuss
the numerical results and draw our conclusions.

\section{Mathematical Formalism}

\label{sect2} Following the general approach of ref.\cite{21}, we present a
deformed $FRW$ solution in a non commutative ($NCG$) gauge gravity where the
structure of the space-time is affected by the commutation relations: 
\begin{equation}
\left[ \widehat{x}^{\mu },\widehat{x}^{\nu }\right] =i\Theta^{\mu \nu }
\,\left( \hbar =c=1\right)
\end{equation}
where $\Theta^{\mu \nu }$ $\left( \mu ,\nu =\overline{1,4}\right) $ are
antisymmetric canonical parameters. The $NCG$ gauge fields (spin connection)
are denoted by $\widehat{\omega }_{\mu }^{AB}\left( x,\Theta^{\mu \nu
}\right)$ and can be expanded in power series of the $\Theta^{\mu \nu }$ as: 
\begin{equation}
\widehat{\omega }_{\mu }^{AB}=\omega _{\mu }^{AB}-i\Theta ^{\upsilon \rho
}\omega _{\mu \nu \rho }^{AB}+\Theta ^{\upsilon \rho }\Theta ^{\lambda \tau
}\omega _{\mu \upsilon \rho \lambda \tau }^{AB}+...  \label{2}
\end{equation}
where 
\begin{equation}
\omega _{\mu \nu \rho }^{AB}=\frac{1}{4}\left\{ \omega _{\nu },\partial
_{\rho }\omega _{\mu }+R_{\rho \mu }\right\} ^{AB}
\end{equation}
\begin{equation}
\omega _{\mu \upsilon \rho \lambda \tau }^{AB}=\frac{1}{32}\left[ 
\begin{array}{c}
\left\{ \omega _{\lambda },\partial _{\tau }\left\{ \omega _{\lambda
},\partial _{\rho }\omega _{\mu }+R_{\rho \mu }\right\} \right\} +2\left\{
\omega _{\lambda },\left\{ R_{\tau \nu },R_{\mu \rho }\right\} \right\} \\ 
-\left\{ \omega _{\lambda },\left\{ \omega _{\nu },D_{\rho }R_{\tau \mu
}+\partial _{\rho }R_{\tau \mu }\right\} \right\} -\left\{ \left\{ \omega
_{\nu },\partial _{\rho }\omega _{\lambda }+R_{\rho \lambda }\right\}
,\partial _{\tau }\omega _{\mu }+R_{\tau \mu }\right\} \\ 
+2\left[ \partial _{\nu }\omega _{\lambda },\partial _{\rho }\left( \partial
_{\tau }\omega _{\mu }+R_{\tau \mu }\right) \right]%
\end{array}
\right] ^{AB}
\end{equation}
with 
\begin{equation}
\left\{ \alpha ,\beta \right\} ^{AB}=\alpha ^{AC}\beta _{C}^{B}+\beta
^{AC}\alpha _{C}^{B}
\end{equation}
and the covariant derivative $D_{\mu }R_{\rho \sigma }^{AB}$ is such that:

\begin{equation}
D_{\mu }R_{\rho \sigma }^{AB}=\partial _{\mu }R_{\rho \sigma }^{AB}+\left(
\omega _{\mu }^{AC}R_{\rho \sigma }^{DB}+\omega _{\mu }^{BC}R_{\rho \sigma
}^{DA}\right) \eta _{CD}
\end{equation}
here 
\begin{equation}
R_{\mu \nu }^{AB}=e^{A\rho }e^{B\sigma }R_{\mu \nu \rho \sigma }
\end{equation}
where $e^{A\rho },$ $R_{\mu \nu \rho \sigma }$ and $\omega _{\mu }^{AB}$ are
the ordinary commutative inverses of the vierbein (tetrad), Riemanian tensor
and spin connection respectively (Greek and Latin indices are for curved and
flat space-time respectively).

Now, if we assume a vanishing commutative torsion, the components of the $%
NCG $ tetrad fields $\widehat{e}_{\mu }^{A}$ read:

\begin{equation}
\widehat{e}_{\mu }^{A}=e_{\mu }^{A}-i\Theta ^{\nu \rho }e_{\mu \nu \rho
}^{A}+\Theta ^{\nu \rho }\Theta ^{\lambda \tau }e_{\mu \nu \rho \lambda \tau
}^{A}+....  \label{8}
\end{equation}
where

\begin{equation}
e_{\mu \nu \rho }^{A}=1/4\left\{ \omega _{\nu }^{AC}\partial _{\rho }e_{\mu
}^{D}+\left( \partial _{\rho }\omega _{\mu }^{AC}+R_{\rho \mu }^{AC}\right)
e_{\nu }^{D}\right\} \eta _{CD}
\end{equation}
and

\begin{equation}
e_{\mu \nu \rho \lambda \tau }^{A}=1/32\left[ 
\begin{array}{c}
-\omega _{\lambda }^{AB}\left( D_{\rho }R_{\tau \mu }^{CD}+\partial _{\rho
}R_{\tau \mu }^{CD}\right) e_{\nu }^{m}\eta _{Dm}-\left\{ \omega _{\nu
},\partial _{\rho }\omega _{\lambda }+R_{\rho \lambda }\right\}
^{AB}\partial _{\tau }e_{\mu }^{C} \\ 
-\left\{ \omega _{\nu },D_{\rho }R_{\tau \mu }+\partial _{\rho }R_{\tau \mu
}\right\} ^{AB}e_{\lambda }^{C}-\partial _{\tau }\left\{ \omega _{\nu
},\partial _{\rho }\omega _{\mu }+R_{\rho \mu }\right\} ^{AB}e_{\lambda }^{C}
\\ 
-\omega _{\lambda }^{AB}\partial _{\tau }(\omega _{\nu }^{CD}\partial _{\rho
}e_{\mu }^{m}+\left( \partial _{\rho }\omega _{\mu }^{CD}+R_{\rho \mu
}^{CD}\right) e_{\nu }^{m})\eta _{Dm}+2\partial _{\nu }\omega _{\lambda
}^{AB}\partial _{\rho }\partial _{\tau }e_{\mu }^{C} \\ 
-2\partial _{\rho }\left( \partial _{\tau }\omega _{\mu }^{AB}+R_{\tau \mu
}^{AB}\right) \partial _{\nu }e_{\lambda }^{C}+2\left\{ R_{\tau \nu },R_{\mu
\rho }\right\} ^{AB}e_{\lambda }^{C} \\ 
+\left( \partial _{\tau }\omega _{\mu }^{AB}+R_{\tau \mu }^{AB}\right)
\left( \omega _{\nu }^{CD}\partial _{\rho }e_{\lambda }^{m}+(\partial _{\rho
}\omega _{\lambda }^{CD}+R_{\rho \mu }^{CD}\right) e_{\nu }^{m})\eta _{Dm}%
\end{array}
\right] \eta _{BC}
\end{equation}
a real $NCG$ metric $\widehat{g}_{\mu \nu }$ is given by \cite{22,23}:  
%%%%%%%%%%%%%%%%%%%%%%%%%%%%%%%%%%%%%%%%%%%%%%%%%%%%%%%%%%
\begin{equation}
\widehat{g}_{\mu \nu }=(1/2)\eta _{AB}\left( \widehat{e}_{\mu }^{A}\ast 
\widehat{e}_{\nu }^{B+}+\widehat{e}_{\nu }^{B+}\ast \widehat{e}_{\mu
}^{A}\right)  \label{11}
\end{equation}
here the superscript \textquotedblleft $+$\textquotedblright\ denotes the
complex conjugate and \textquotedblleft $\ast $\textquotedblright\ the
Moyal-Weyl star product defined as: 
\begin{equation}
f(x)\ast g(y)=f(x)\exp \left[ \frac{i}{2}\Theta ^{\mu \nu }\frac{ 
\overleftarrow{\partial }}{\partial x^{\mu }}\frac{\overrightarrow{\partial }
}{\partial y^{\nu }}\right] g(y)\mid _{x=y}
\end{equation}
In what follows, we take the signature of space-time $\left( +,+,+,-\right)$%
, and in order to simplify our calculations, we choose the component $%
\Theta^{12}$ as the only non vanishing space-space components. It is worth
to mention as it was pointed out in ref.\cite{24} a non vanishing time-space
components $\Theta^{i4}$ $\left( i=\overline{1,3} \right) $ will affect
unitarity and causality\cite{24,25,26}.Now, let us start from a spherically,
isotropic and homogeneous flat space $(k=0)$ $FRW$ universe. To keep our
results more transparent and simple, we use dimensionless spherical
coordinates $\widehat{r}=\frac{r}{r_{0}},\theta ,\varphi $ and time $%
\widehat{t}=\frac{t}{t_{0}}$ ( $r_{0}$ and $t_{0}$ are some cosmological
scales parameters which will be specified later), and a power law formula
for the scale factor $a\left( \widehat{t}\right) $ of the form $a\left(%
\widehat{t}\right) =\widehat{t}^{\beta }$. Because all the measured proper
distances $r$ between co-moving points increase proportionally to $a(t)$,
then one can write $\widehat{r}\approx \frac{a(t)}{a(t_{0})}=\widehat{t}
^{\beta }$. Moreover, the time scale $t_{0}$ is related to the parameter $%
\beta $ and matter density $\rho _{0}$ through the relation (see Appendix {%
\ref{AppenA}}): 
\begin{equation}
t_{0}=\beta \left( \frac{3}{\chi \rho _{0}}\right) ^{\frac{1}{2}}
\end{equation}
where $\chi =\frac{8\pi G}{c^{4}}$ and $G$ the Newton's gravitational
constant.

Now, straightforward calculations using a Maple package, and writing $%
\Theta^{12}=\widetilde{\Theta }$ ($\widetilde{\Theta }$ =$\frac{\Theta}{
\Lambda }$ and $\Lambda $ is an $NCG$ scale factor), give the following
non-zero components of the deformed metric $\widehat{g}_{\mu \nu }$ up to
the $O\left( \Theta^{2}\right) $ as it is shown in Appendix {\ref{AppenB}}.

Notice that $\widehat{g}_{\mu \nu }$ is real and in general not symmetric.
Moreover, $NCG$ has generated a non homogeneous and anisotropic universe,
the corresponding components of the tetrad are in general complex (see
Appendix {\ref{AppenB}}). Similarly, direct but tedious calculations give
the non vanishing $NCG$ spin connection $\widehat{\omega }_{\mu }^{AB}$ up
to the $O\left( \Theta^{2}\right) $ (see Appendix {\ref{AppenB}}).

Notice that contrary to the ordinary commutative case, $\widehat{\omega }
_{\mu }^{AB}$ is in general a complex and not completely antisymmetric with
respect to $A$ and $B$.

\section{NCG Neutrino-Antineutrino Asymmetry}

\label{sect3}

In an $NCG$ curved space-time, the generalized Dirac Lagrangian density $%
\mathcal{L}$\ is assumed to have the form \cite{7}, \cite{14} : 
\begin{equation}
{\mathcal{L}} =\sqrt{-\widehat{g}}\ast \widehat{{\Psi }}\ast \left( -i\gamma
^{\mu }\ast \widehat{D}_{\mu }\ast -m\right) \widehat{\Psi }+c.c
\end{equation}
where the $NCG$ covariant derivative $\widehat{D}_{\mu }$ is given by: 
\begin{equation}
\widehat{D}_{\mu }=\partial_{\mu }-\frac{i}{4}\widehat{\omega }%
_{\mu}^{AB}\Sigma_{AB}
\end{equation}
with 
\begin{equation}
\gamma^{\mu }=\widehat{e}_{a}^{\mu }\gamma ^{a},\, \Sigma\_{\ \left[ AB%
\right] }=\frac{i}{2}\left[ \gamma _{A},\gamma _{B}\right] ,\,
\Sigma_{\left( AB\right) }=\frac{1}{2}\left\{ \gamma _{A},\gamma
_{B}\right\} =\eta _{AB}
\end{equation}
($\gamma _{i}^{\prime }$ s are the Dirac Gamma matrices in the flat
Minkowski space-time and $\widehat{\Psi }$ the $NCG$ $4$-components Dirac
spinor). Here \textquotedblleft $c.c$\textquotedblright\ stands for complex
conjugate and $\widehat{e}_{a}^{\mu }$ the inverse of $NCG$ vierbein. Using
the fact that 
\begin{equation}
\frac{1}{2}\left[ \gamma _{d}\Sigma_{\left[ AB\right] }+\Sigma_{\left[ AB%
\right] }\gamma _{d}\right] =\varepsilon _{fdab}\gamma ^{e}\gamma ^{5}
\end{equation}
and 
\begin{equation}
\frac{i}{2}\left[ \gamma _{d}\Sigma_{\left[ AB\right]} -\Sigma_{\left[ AB%
\right]} \gamma_{d}\right] =g_{db}\gamma_{a}-g_{da}\gamma_{b}
\end{equation}
as well as the $NCG$ orthogonality relation 
\begin{equation}
\widehat{e}_{a}^{\mu }\ast \widehat{e}_{\mu b}=\delta_{ab}
\end{equation}
where $\varepsilon_{fdab}$ is the $4$-rank totally antisymmetric tensor, one
can show that the corresponding $NCG$ Dirac equation takes the form: 
\begin{equation}
\left[ \gamma^{f}\left( i\partial_{f}+\widehat{A}_{f}\right) +\gamma
^{f}\gamma^{5}\widehat{B}_{f}\right] \ast \widehat{\Psi }=0
\end{equation}
where 
\begin{eqnarray}
\partial _{f} &=&\widehat{e}_{f}^{\mu }\partial _{\mu } \\
A^{f} &=&\Im(\widehat{e}^{\mu f}\sum_{a=1}^{4}\omega _{\mu }^{aa})+%
\Re\left( \widehat{e}_{d}^{\mu }\left( \widehat{\omega }_{\mu }^{fd}- 
\widehat{\omega }_{\mu }^{df}\right) \right) +O\left( \Theta ^{3}\right) \\
B_{f} &=&\left[ \Im\left( \widehat{e}^{\mu d}\widehat{\omega }_{\mu
}^{ab}\right) +\frac{1}{4}\Theta ^{\rho \sigma }\Theta ^{\alpha \beta
}\left( \partial _{\rho }\partial _{\alpha }e^{\mu d}\right) \left( \partial
_{\sigma }\partial _{\beta }\omega _{\mu }^{ab}\right) \right] \varepsilon
_{fdab}+O\left( \Theta ^{3}\right)
\end{eqnarray}
one can show (as in ref.\cite{27}) that the dispersion relations for the
left and right chiral fields (particles $X$ and antiparticles $\overline{X}$
) read:

\begin{equation}
E_{X}=\sqrt{\left( \overrightarrow{P}-\overrightarrow{\Lambda }^{X}\right)
^{2}+m^{2}}+\Lambda_{4}^{X}
\end{equation}
and 
\begin{equation}
E_{\overline{X}}=\sqrt{\left( \overrightarrow{P}-\overrightarrow{\Lambda }^{ 
\overline{X}}\right) ^{2}+m^{2}}+\Lambda _{4}^{\overline{X}}
\end{equation}
where 
\begin{equation}
\overrightarrow{\Lambda }^{X}=\overrightarrow{B}+\overrightarrow{A}, \,\, \,
\, \, \, \, \, \, \, \, \, \, \, \, \overrightarrow{\Lambda }^{\overline{X}%
}=- \overrightarrow{B}+\overrightarrow{A}
\end{equation}
and 
\begin{equation}
\Lambda _{4}^{X}=B_{4}+A_{4} \, \, \, \, \, \, \, \, \, \, \, \, \Lambda
_{4}^{ \overline{X}}=-B_{4}+A_{4}
\end{equation}
In fact, this result was expected because of $CPT$ violation which affects
the dispersion relations of particles and antiparticles leading to a
difference between their energies. In our case, the violation of $CPT$ can
be induced by:

\begin{itemize}
\item[1)] Curvature in a certain non trivial anisotropic $NCG$ geometry. In
fact, fermions in a curved space-time can interact via an axial vector
current due to their spin with space-time curvature or torsion \cite{27}.
Therefore, the space-time curvature background\ has the effect of inducing
an axial vector field especially in certain anisotropic space-time
geometries like in $NCG$.

\item[2)] $NCG$ : since it violates Lorentz invariance leading to a non
conservation of $CPT$.
\end{itemize}

Thus, in our case the violation of $CPT$ is induced by both the curvature
and non commutativity of the space-time.

Now, in order to derive the neutrino-antineutrino asymmetry in the context
of $NCG$, one has to have a thermal equilibrium background coming after the
neutrino decoupling. In fact, the synthetics of light elements depends
strongly on the ratio of number of neutrinos / number of protons freezout
abundance determined by the interplay between the weak interaction and
expansion rate of the universe. Both are influenced by the neutrino
decoupling temperature. The neutrino decoupling means that the neutrinos do
not interact with baryonic matter and consequently do not influence the
dynamics of the universe at early stage. The decoupling happens when the
weak interaction rate of neutrinos is smaller than the expansion rate of the
universe ( $kT_{decoup}^{\nu }\approx 1 MeV,$ $t_{decoup}\approx 1 seconde).$
After decoupling, a thermal equilibrium background of relativistic neutrinos
is expected with an effective temperature at late time $T^{\nu }\approx 0.71
T_{\gamma }.$ Now, at this equilibrium temperature and using the Fermi-Dirac
distribution together with the result:

\begin{equation}
\int_{0}^{\infty }dx\frac{x^{m}}{1+z^{-1}e^{x}}=z\Gamma \left( m+1\right)
\Phi \left( -z,m+1,1\right)
\end{equation}
where $\Gamma \left( x\right) $ is the Euler Gamma function and $\Phi \left(
z,s,a\right) $ the Lerch transcendent function given by 
\begin{equation}
\Phi \left( z,s,\alpha \right) =\sum_{m=0}^{\infty }\frac{z^{n}}{\left(
n+\alpha \right)^s }
\end{equation}
the analytical expression of the difference between the neutrino and
antineutrino $NCG$ number density $\Delta n_{NCG}$ is given by (in the
system where the Boltzman constant $k=1$): 
\begin{equation}
\Delta n_{NCG}\approx\frac{\left({T_\nu}\right)^3}{2 \pi^2}
\sum_{l=1}^{\infty}\frac{(-1)^l}{l^3} \sinh\left(\frac{l \widehat{B}_4}{T_\nu%
}\right)  \label{24}
\end{equation}
where $\widehat{B}_4$ has as an expression  
\begin{equation}
\widehat{B_4}=\frac{\Theta}{\Lambda}t_0^2 \left[\frac{\hat{t}^{-\beta-1}}{2
\sin\theta}-\frac{1}{4}\beta^2 \sin \theta \,\hat{t}^{2\beta-2}\right]
\label{25}
\end{equation}
here $t_0$ stands for the decoupling time $t_{decoup}$.

Notice that since $\widehat{B_4}$ is propotional to $\Theta$, as a first
approximation, eq.(\ref{24}) reduces to 
\begin{equation}
\bigtriangleup n_{_{NCG}}\approx \frac{(T_{\nu })^{2}}{2\pi ^{2}}\widehat{B}%
_{4}\zeta\left( 2\right)
\end{equation}
where $\zeta\left( 2\right) $ is the Dirichlet zeta function. Now, using the 
$CMB$ photon number density $n_{\gamma }$, the ratio $R_{1}=$\ $\frac{%
\bigtriangleup n_{_{_{NCG}}}}{n_{\gamma }}$ reads: 
\begin{equation}
R_{1}=\frac{\pi ^{2}}{12\zeta(3)}\left( \frac{\widehat{B}_{4}}{T_{\nu }}%
\right) \left( \frac{T_{\nu }}{T_{\gamma }}\right) ^{3}  \label{27}
\end{equation}
where $T_{\gamma }$ is the $CMB$ \ photon temperature. Now, it is clear from
eq.(\ref{27}) that contrary to the commutative $FRW$ universe (with
isotropy, homogeneity and space-time spherical symmetry) where there is no
geometrical contribution to the neutrino-antineutrino asymmetry $\Delta n$,
the $NCG$ space-time has generated anisotropy, non homogeneity and  Broken
spherical symmetry leading to a dynamical non vanishing net asymmetry $%
\Delta n_{NCG}$ which has a non commutative geometrical origin.

\section{Discussions and Conclusions}

\label{sect4}

We have found that if neutrinos are propagating in a curved $NCG$ universe,
where a space-time anisotropy as well as Lorentz violation invariance are
generated, a net asymmetry between neutrinos and antineutrinos arises at the
thermodynamical equilibrium. Thus, we have shown that a new source of a
leptonic and antileptonic asymmetry which has as an origin the
noncommutative geometrical structure of space-time can be generated leading
to a new mechanism explaining matter-antimatter asymmetry. It is worth to
mention that in the standard cosmology at present time, the ratio $R_{2}=%
\frac{\Delta n}{n_{\gamma }}$ is given by \cite{1},\cite{28} 
\begin{equation}
R_{2}= \frac{\pi ^{2}}{12\zeta(3)}\left( \frac{T_{\nu }}{T_{\gamma }}
\right) ^{3}\left( \frac{\mu _{\nu }}{T_{\nu }}\right)
\end{equation}
$\mu _{\nu }$ is the neutrino chemical potential. At the decoupling time
where $t_{0}\approx 1s$, $\frac{T_{\nu }}{T_{\gamma }} \approx 0.71$ and $%
\left\vert \frac{\mu _{\nu }}{T_{\nu }}\right\vert \lesssim 0.04,$ the ratio 
$R_{2}\lesssim 9\times 10^{-3}$. Now, if we consider the $\theta$-averaged $%
\left\langle \widehat{B}_{4}\right\rangle _{\theta }$ $\left( -\frac{\pi }{2}%
\leq \theta \leq \frac{\pi }{2}\right) $ at present time with the leading
term $\sim \widehat{t}^{2\beta -2},$ $R_{1}$ takes the simple form: 
\begin{equation}
R_{1}\mid _{t=t_{0}}\approx \frac{\pi ^{2}}{12\zeta(3)}\left[ -\frac{%
\widetilde{ \Theta }}{4 t_{0}^{2}}\beta ^{2}\frac{\widehat{t}^{2\beta -2}}{%
T_{\nu }}\right] \left( \frac{T_{\nu }}{T_{\gamma }}\right) ^{3},\,
\end{equation}
here $\widehat{t}=\frac{t_{0b}}{t_{0}}$, using the fact that $kT_{\nu
}\approx 2\times 10^{-23}J$ and $t_{0b}\approx 0.3\times 10^{18}s,$ one
gets: 
\begin{equation}
\left\vert R\right\vert =\left\vert \frac{R_{1}}{R_{2}}\right\vert
_{t=t_{0b}}\approx \frac{\widetilde{\Theta }\beta ^{2}}{12.8}\times
10^{-45}\left( 3\times 10^{17}\right) ^{2\beta -2}J
\end{equation}
which means that if $\left\vert R\right\vert \sim O(1),$ then $\Lambda \sim 
\frac{ \Theta \beta ^{2}}{20,5}\times 10^{-38}\left( 3\times 10^{17}\right)
^{2\beta -2}TeV.$

Notice that the $NCG$ scale $\Lambda $ and $\beta =\frac{2}{3\left( 1+\omega
\right) }$ are intimately related. Furthermore, for an accelerated expansion
of the universe one has $\beta > 1$ and therefore: $\Lambda \succ \frac{%
\Theta }{20,5} \times 10^{-38}TeV.$ This is a new lower bound of the non
commutative scale $\Lambda $ $\left( \Theta < 1\right) .$

\begin{figure}[h]
\centerline{\includegraphics[width=6.5cm]{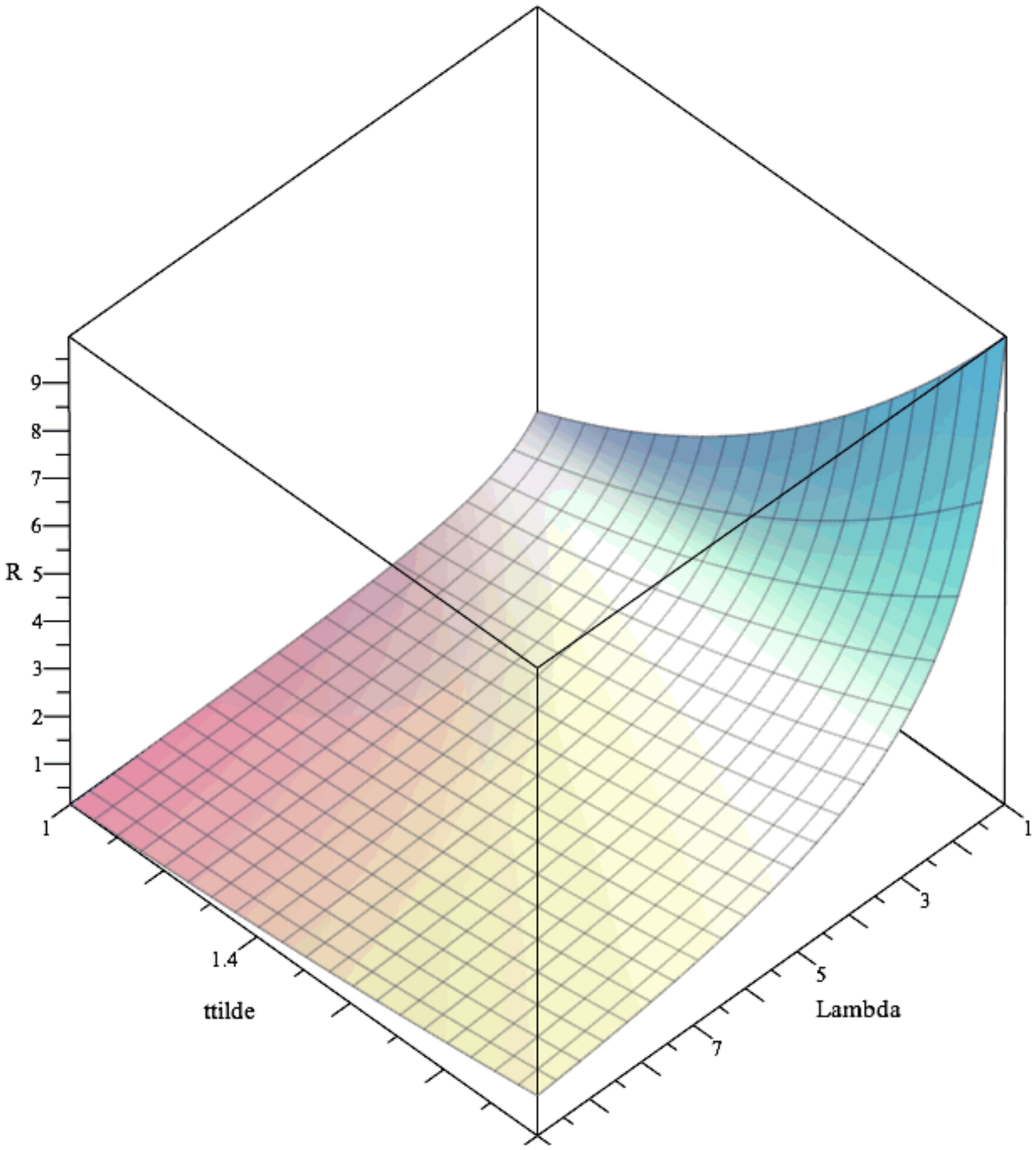}}
\caption{Ratio $R$ as a function of $\frac{\Lambda }{\Theta }$ and $%
\widetilde{t}$ for $\protect\beta \approx 2.38.$.}
\label{f1}
\end{figure}
Fig.{\ref{f1}}. shows the ratio $R$ as a function of the $NCG$ parameter $%
\frac{ \Lambda }{\Theta }$ ( denoted by Lambda and the reduced time in a $TeV
$ unit), the reduced time $\widetilde{t}$ (denoted by ttilde, $\widehat{t}
=10^{17}\widetilde{t})$ for a fixed $\beta \approx 2.38$ or equivalently the
equation of state parameter ($EOS$) parameter $\omega \approx -0.75$. Notice
that $R$ is an increasing function of time for a fixed $\frac{\Lambda }{%
\Theta }.$ Moreover, for smaller values of $\frac{\Lambda }{\Theta }$ of $%
O(1TeV)$, $R\sim 10$ when $\widehat{t}\sim 2t_{ob}$ ($t_{ob}$ is the
observed time). This shows that $R$ becomes important and $NCG$ effects
dominate in comparison to the one of the standard cosmology. It is worth to
notice that if the $NCG$ effects become relevant ($R > 1),$ one gets an
upper limit for $\frac{\Lambda }{\Theta }$ of $O(1\, TeV)$.

\begin{figure}[h]
\centerline{\includegraphics[width=6.5cm]{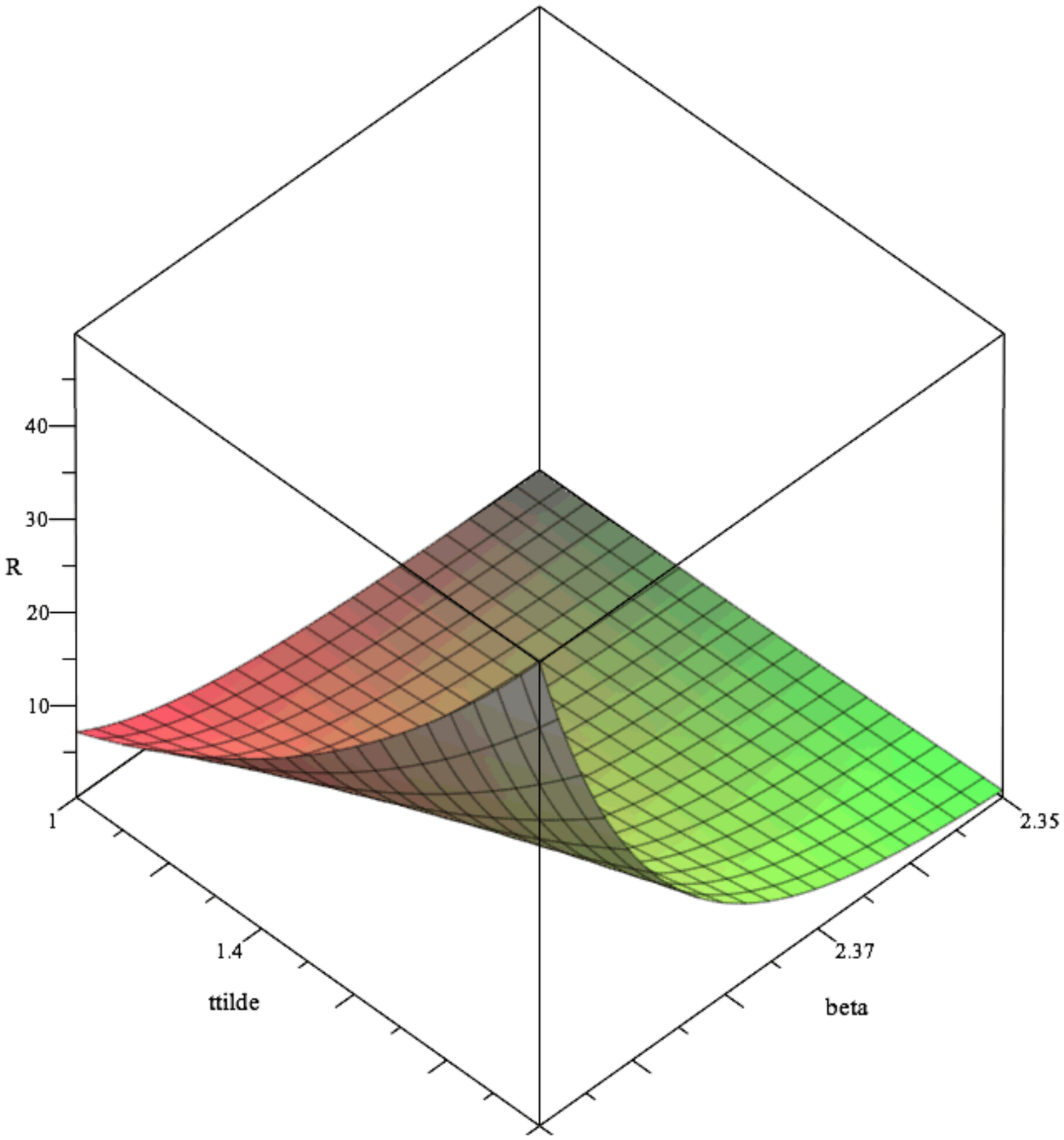}} 
\caption{Ratio $R$ as a function of $\protect\beta $ and $\widetilde{t}$ for 
$\frac{\Lambda }{\Theta }\approx 1$ $TeV.$}
\label{f2}
\end{figure}
Fig.{\ref{f2}}. displays the ratio $R$ as a function of the $\beta$
parameter (denoted by beta) and reduced time $\widetilde{t}$ for a fixed $NCG
$ parameter $\frac{\Lambda }{ \Theta }\sim 1$ $TeV.$ Notice that the ratio $R
$ is very sensitive to the variations of $\beta $; e.g. if $\beta \sim
2.35-2.4$, and $\widetilde{t}\sim 1-2$, the ratio $R\sim 1-50.$ 
\begin{figure}[h]
\centerline{\includegraphics[width=6.5cm]{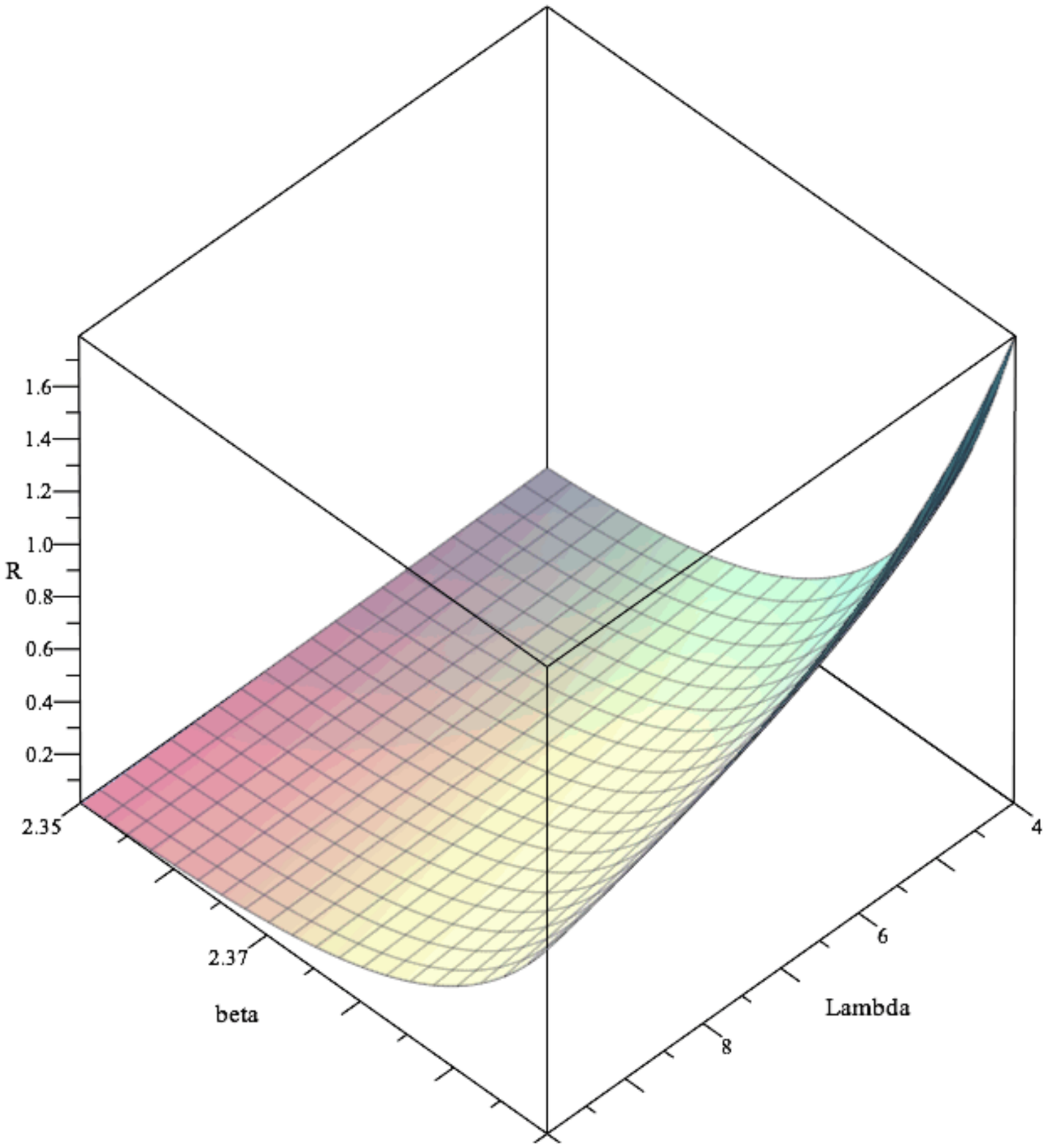}} 
\caption{Ratio $R$ as a function of $\frac{\Lambda }{\Theta }$ and $\protect%
\beta $ at present time.}
\label{f3}
\end{figure}

Fig.{\ref{f3}}. shows the ratio $R$ as a function of $\beta$ and $\frac{%
\Lambda }{ \Theta }$ at the present time. For $\frac{\Lambda }{\Theta }\sim 4
$ $TeV$ and $\beta \sim 2,$ the ratio$\ R$ reaches the value $1.8.$ 
\begin{figure}[h]
\centerline{\includegraphics[width=6.5cm]{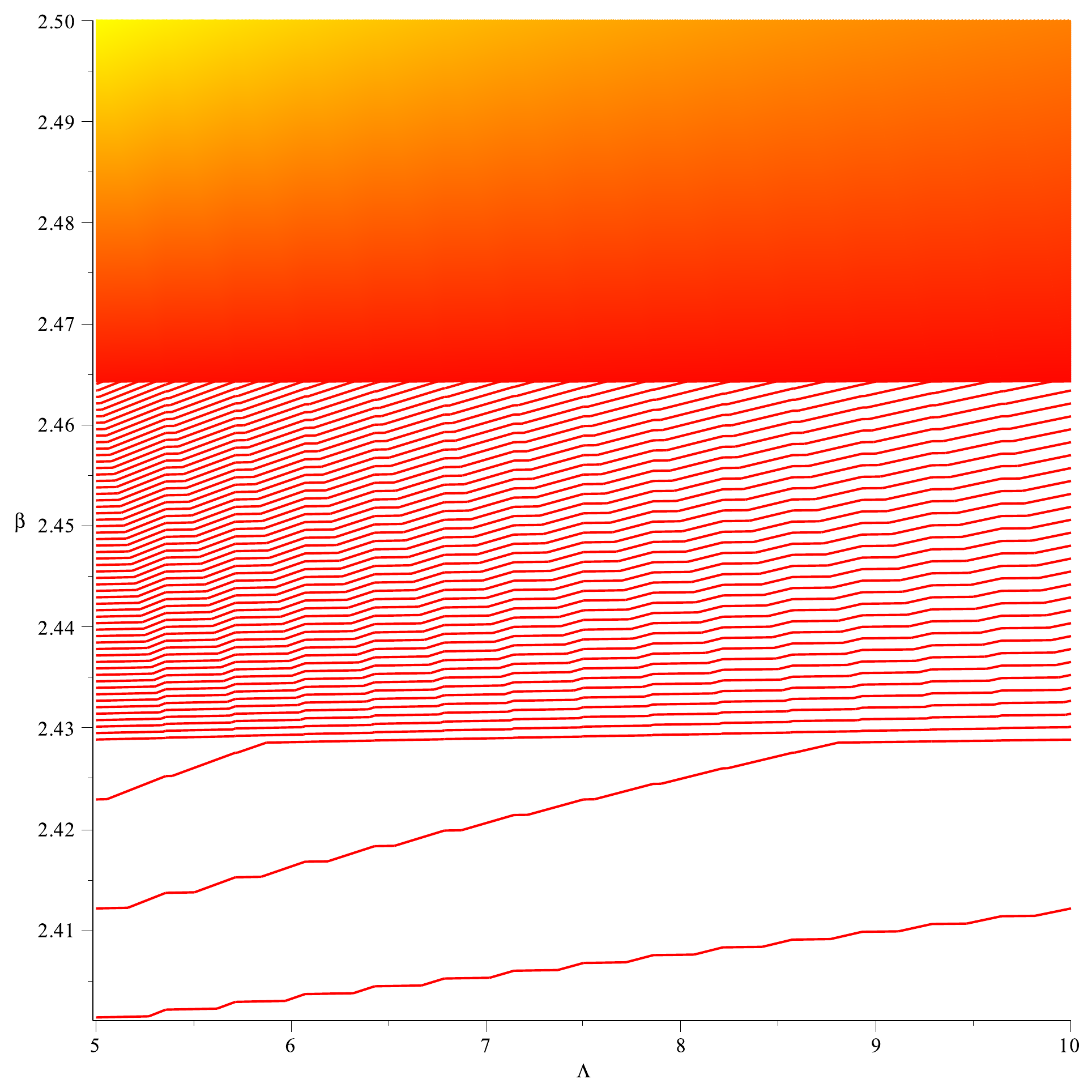}} 
\caption{Contour plots $\left( \frac{\Lambda }{\Theta },\protect\beta %
\right) $ at present time for a fixed $R.$}
\label{f4}
\end{figure}

Fig.{\ref{f4}}. displays the contour plots $\left( \frac{\Lambda(TeV) }{%
\Theta },\beta \right) $ for a fixed $R$ at the present time. Notice that
for a fixed $\beta $ or $\omega $ as $R$ increases; the value of the $NCG$
parameter $\frac{\Lambda }{\Theta }$ decreases and becomes relevant at a $TeV
$ scale. As an important and novel results, the $NCG$ $\frac{\Lambda }{%
\Theta }$ parameter is strongly related to the neutrino-antineutrino
asymmetry which has as an origin the space-time $NCG$ structure, and the $EOS
$ parameter $\omega $ (or equivalently $\beta $)$.$ Moreover, this asymmetry
is dynamical in the sense that it is a time dependent quantity. It increases
with time for an accelerated expansion of the universe. Finally, as a
conclusion a new mechanism from the non commutativity of space-time
generating the matter-antimatter asymmetry was found.

\section{Acknowledgments}

We are very grateful to the Algerian Ministry of higher education and
scientific research and the $DGRSDT$ for the financial support. \appendix

\section{Appendix}

\label{AppenA}

From Friedman's equations: 
\begin{equation}
2\frac{\overset{..}{a}}{a}+\chi \left( P+\frac{1}{3}\rho \right) =0
\label{A-1}
\end{equation}
and 
\begin{equation}
2\frac{\overset{..}{a}}{a}+\chi \left( P+\rho \right) -2\frac{\overset{.}{a}
^{2}}{a}=0  \label{A-2}
\end{equation}
($\rho $ and $P$ are the matter density and pressure respectively), one can
get the following continuity equation (in what follows, we use the system
where $\chi =1$ and "$.$" stands for time derivative) 
\begin{equation}
\overset{.}{\rho }+3\frac{\overset{.}{a}}{a}\left( P+\rho \right) =0
\label{A-3}
\end{equation}
leading to: 
\begin{equation}
\frac{d}{dt}\left( \rho a^{3}\right) +\omega \rho \frac{da^{3}}{dt}=0
\label{A-4}
\end{equation}
for a perfect fluid where $P=\omega \rho .$ One can show that the solution
of eq.(\ref{A-4}) is of the form $\rho =D\, a^{-3\left( 1+\omega \right) }.$
Moreover, from eqs.(\ref{A-1}) and (\ref{A-2}), one can obtain: 
\begin{equation}
\overset{.}{a}^{2}=\rho \frac{a^{2}}{3}  \label{A-5}
\end{equation}
consequently,

\begin{equation}
a(t)=\left( \frac{3}{2}\left( 1+\omega \right) \right) ^{\frac{2}{3\left(
1+\omega \right) }}\left( \frac{\chi D}{3}\right) ^{^{\frac{1}{3\left(
1+\omega \right) }}}t^{\frac{2}{3\left( 1+\omega \right) }}  \label{A-6}
\end{equation}
Now, it is clear that: 
\begin{equation}
\beta =\frac{2}{3\left( 1+\omega \right) }  \label{A-7}
\end{equation}
\begin{equation}
D =\rho _{0}a_{0}^{3\left( 1+\omega \right) }  \label{A-8}
\end{equation}
\begin{equation}
t_{0} =\beta \left( \frac{3}{\chi \rho _{0}}\right) ^{\frac{1}{2}}
\label{A-9}
\end{equation}

\section{Appendix}

\label{AppenB}

Starting from eqs.(\ref{2}), (\ref{8}) and (\ref{11}) and using maple $18$
package, one obtains the following complex $NCG$ vierbeins, the metric
components and the spin connection components respectively; (We set : $h=%
\frac{r_{0}}{t_{0}}$ and $\widetilde{\Theta }=\frac{ \Theta }{\Lambda t_{0}})
$, in what follow, the upper (respectively lower) indices stand for flat
(respectively curved) space-time,

\begin{equation}
\widehat{e}_{1}^{1}=\frac{\widehat{t}^{\beta }}{128}\left[ 128+\frac{%
\widetilde{\Theta }^{2}}{12}\beta^{2}\widehat{r}^{2}\widehat{t}^{3\beta-2}
h^{2}\left( 14\beta^{2}\widehat{r}^{2} \widehat{t}^{-2}\sin^{2}\theta +14
\sin ^{2}\theta +1 \right) \right]  \label{B-1}
\end{equation}

\begin{equation}
\widehat{e}_{2}^{1}=-\frac{3}{256}\widetilde{\Theta }^{2}\widehat{r}r_{0} 
\widehat{t}^{\beta }\sin 2\theta \left( 4-\widehat{t}^{2\beta -2}\beta ^{2} 
\widehat{r}^{2}h^{2}\right)  \label{B-2}
\end{equation}
\begin{equation}
\widehat{e}_{3}^{1}=-\frac{i\widetilde{\Theta }}{2}\widehat{r}r_{0}\widehat{%
t }^{\beta }  \label{B-3}
\end{equation}
\begin{equation}
\widehat{e}_{4}^{1}=-\frac{1}{8}\widetilde{\Theta }^{2}\beta ^{2}\left(
\beta -1\right) \widehat{r}^{3}\widehat{t}^{3\beta -3}h^{3}  \label{B-4}
\end{equation}
\begin{equation}
\widehat{e}_{1}^{2}=-\frac{7}{64}\widetilde{\Theta }^{2}\beta ^{2}\widehat{r}
^{2}\widehat{t}^{3\beta -2}h^{2}\sin 2\theta  \label{B-5}
\end{equation}
\begin{equation}
\widehat{e}_{2}^{2}=-\frac{1}{128}\widehat{r}r_{0}\widehat{t}^{\beta
}\left\{ -128+\widetilde{\Theta }^{2}\widehat{t}^{\beta }\left[ 
\begin{array}{c}
\beta ^{2}\widehat{r}^{2}\widehat{t}^{2\beta -1}h^{2}\left( -7\sin
^{2}\theta +18\cos ^{2}\theta \right) \\ 
-17\beta ^{4}\widehat{r}^{4}\widehat{t}^{2\beta -4}h^{4}\sin ^{2}\theta - 
\widehat{t}^{\beta }\left( 2\sin ^{2}\theta +1\right)%
\end{array}
\right] \right\}  \label{B-6}
\end{equation}
\begin{equation}
\widehat{e}_{3}^{2}=-\frac{i}{4}\widetilde{\Theta }\widehat{r}r_{0}\widehat{%
t }^{\beta }\left[ \cos 2\theta -\beta ^{2}\widehat{r}^{2}\widehat{t}%
^{2\beta -2}h^{2}\sin ^{2}\theta \right]  \label{B-7}
\end{equation}

\begin{equation}
\widehat{e}_{4}^{2}=-\frac{7}{64}\widetilde{\Theta }^{2}\beta ^{2}\left(
\beta -1\right) \widehat{r}^{3}\widehat{t}^{3\beta -3}h^{3}\sin 2\theta
\label{B-8}
\end{equation}

\begin{equation}
\widehat{e}_{1}^{3}=\widehat{e}_{4}^{3}=0  \label{B-9}
\end{equation}

\begin{equation}
\widehat{e}_{2}^{3}=-\frac{i}{4}\widetilde{\Theta }\beta ^{2}\widehat{r}
^{3}r_{0}\widehat{t}^{3\beta -2}h^{2}\sin \theta  \label{B-10}
\end{equation}

\begin{equation}
\widehat{e}_{3}^{3}=\frac{1}{128}\widehat{r}r_{0}\widehat{t}^{\beta }\sin{%
\theta}\left\{ 128+\widetilde{\Theta }^{2}\left[ 
\begin{array}{c}
-\beta ^{2}\widehat{r}^{2}\widehat{t}^{2\beta -2}h^{2}\left( 5+16\cos
^{2}\theta \right) \\ 
+17\beta ^{4}\widehat{r}^{4}\widehat{t}^{4\beta -4}h^{4}\sin ^{2}\theta +2%
\end{array}
\right] \right\}  \label{B-11}
\end{equation}

\begin{equation}
\widehat{e}_{1}^{4}=\frac{1}{128}\widetilde{\Theta }^{2}\beta \widehat{r} 
\widehat{t}^{2\beta -1}h\left[ 5+\sin ^{2}\theta +22\beta ^{2}\widehat{r}%
^{2} \widehat{t}^{2\beta -2}h^{2}\right]  \label{B-12}
\end{equation}

\begin{equation}
\widehat{e}_{2}^{4}=-\frac{3}{256}\beta \widetilde{\Theta }^{2}\widehat{r}
^{2}r_{0}h\widehat{t}^{2\beta -1}\left[ -4+\beta ^{2}\widehat{r}^{2}\widehat{
t}^{2\beta -2}h^{2}\right]  \label{B-13}
\end{equation}

\begin{equation}
\widehat{e}_{3}^{4}=\frac{i}{4}\widetilde{\Theta }\widehat{r}^{2}r_{0}\beta
h \widehat{t}^{2\beta -1}\sin 2\theta  \label{B-14}
\end{equation}

\begin{equation}
\widehat{e}_{4}^{4}=-\frac{1}{64}\left\{ -64+\widetilde{\Theta }^{2}\left[ 
\begin{array}{c}
-18\beta ^{3}\left( \beta -1\right) \widehat{r}^{4}\widehat{t}^{4\beta
-4}h^{4}\sin ^{2}\theta \\ 
-\widehat{r}^{2}h^{2}\beta \left( \beta -1\right) \widehat{t}^{2\beta
-2}\left( 6\sin ^{2}\theta +1\right)%
\end{array}
\right] \right\}  \label{B-15}
\end{equation}

$\bullet $ $\widehat{g}_{\mu \nu }$ components: 
\begin{equation}
\widehat{g}_{11}=\frac{\widehat{t}^{2\beta }}{64}\left\{ 64+\widetilde{
\Theta }^{2}\beta ^{2}\widehat{t}^{2\beta -2}\widehat{r}^{2}h^{2}\left[
14\left( \beta ^{2}\widehat{t}^{2\beta -2}\widehat{r}^{2}h^{2}-1\right) \sin
^{2}\theta -3\right] \right\}  \label{B-16}
\end{equation}

\begin{equation}
\widehat{g}_{12}=-\frac{\widetilde{\Theta }^{2}}{257}\widehat{r}r_{0} 
\widehat{t}^{2\beta }\sin 2\theta \left[ 12+25\beta ^{2}\widehat{t}^{4\beta
-2}\widehat{r}^{2}h^{2}\right]  \label{B-17}
\end{equation}

\begin{equation}
\widehat{g}_{13}=\widehat{g}_{23}=\widehat{g}_{31}=\widehat{g}_{32}=\widehat{
g}_{34}=\widehat{g}_{43}=0  \label{B-18}
\end{equation}

\begin{equation}
\widehat{g}_{14}=\frac{2}{257}\widetilde{\Theta }^{2}\beta \widehat{r} 
\widehat{t}^{2\beta -1}h\left[ 2\beta \left( 19\beta +8\right) \widehat{t}
^{2\beta -2}\widehat{r}^{2}h^{2}\sin ^{2}\theta -10\cos 2\theta \right]
\label{B-19}
\end{equation}

\begin{equation}
\widehat{g}_{21}=-\frac{\widetilde{\Theta }^{2}}{64}\widehat{r}r_{0}\widehat{
t}^{2\beta }\sin 2\theta \left[ 12+25\beta ^{2}\widehat{t}^{4\beta -2} 
\widehat{r}^{2}h^{2}\right]  \label{B-20}
\end{equation}

\begin{equation}
\widehat{g}_{22}=\frac{1}{64}\widehat{r}^{2}r_{0}^{2}\widehat{t}^{2\beta
}\left\{ -64+\widetilde{\Theta }^{2}\left[ 
\begin{array}{c}
12\sin ^{2}\theta +6+21\beta ^{4}\widehat{r}^{4}\widehat{t}^{4\beta -4} \\ 
\times h^{4}\left( 7\sin ^{2}\theta -18\cos ^{2}\theta \right) \beta ^{2} 
\widehat{r}^{2}h^{2}\widehat{t}^{2\beta -2}%
\end{array}
\right] \right\}  \label{B-21}
\end{equation}

\begin{equation}
\widehat{g}_{24}=-\frac{1}{64}\beta ^{2}r_{0}h^3\widehat{r}^{4}\widehat{t}
^{4\beta -3}\sin ^{2}\theta \left[ 25\beta -28\right]  \label{B-22}
\end{equation}

\begin{equation}
\widehat{g}_{33}=\widehat{r}^{2}r_{0}^{2}\widehat{t}^{2\beta }\left( 1+ 
\widetilde{\Theta }^{2}\sin ^{2}\theta \right)  \label{B-23}
\end{equation}

\begin{equation}
\widehat{g}_{41}=-\frac{1}{64}\widetilde{\Theta }^{2}\beta \widehat{r} 
\widehat{t}^{2\beta -1}h\left[ \left( 19\beta \sin ^{2}\theta -8\right) 
\widehat{t}^{2\beta -2}h^{2}\beta \widehat{r}^{2}-8\sin ^{2}\theta \right]
\label{B-24}
\end{equation}

\begin{equation}
\widehat{g}_{42}=\frac{\widetilde{\Theta }^{2}}{64}\beta ^{2}r_{0}\widehat{r}
^{4}\widehat{t}^{4\beta -3}h^{3}\left[ 25\beta -28\right]  \label{B-25}
\end{equation}

\begin{equation}
\widehat{g}_{44}=\frac{1}{32}\left\{ -32+\widetilde{\Theta }^{2}h^{2}\beta 
\widehat{r}^{2}\widehat{t}^{2\beta -2}\left[ 
\begin{array}{c}
-18\beta ^{2}\left( \beta -1\right) \widehat{t}^{2\beta -2}h^{2}\widehat{r}
^{2}\sin ^{2}\theta \\ 
+\left( \beta -1\right) \left( 6\cos ^{2}\theta -7\right)%
\end{array}
\right] \right\}  \label{B-26}
\end{equation}

$\bullet $ Spin connection components at order of $\Theta $ : 
\begin{equation}
\widehat{\omega }_{1}^{12}=\widehat{\omega }_{1}^{13}=\widehat{\omega }
_{1}^{34}=\widehat{\omega }_{1}^{42}=\widehat{\omega }_{1}^{43}=\widehat{
\omega }_{4}^{12}=\widehat{\omega }_{4}^{13}=\widehat{\omega }_{4}^{31}= 
\widehat{\omega }_{4}^{42}=\widehat{\omega }_{4}^{44}=0  \label{B-27}
\end{equation}

\begin{eqnarray}
\widehat{\omega }_{1}^{11} &\simeq &\widehat{\omega }_{1}^{21}\simeq 
\widehat{\omega }_{1}^{22}\simeq \widehat{\omega }_{1}^{24}\simeq \widehat{
\omega }_{1}^{33}\simeq \widehat{\omega }_{1}^{44}\simeq \widehat{\omega }
_{2}^{11}\simeq \widehat{\omega }_{2}^{22}\simeq \widehat{\omega }
_{2}^{33}\simeq \widehat{\omega }_{2}^{41}  \label{B-28} \\
&\simeq &\widehat{\omega }_{2}^{44}\simeq \widehat{\omega }_{4}^{11}\simeq 
\widehat{\omega }_{4}^{14}\simeq \widehat{\omega }_{4}^{21}\simeq \widehat{
\omega }_{4}^{22}\simeq \widehat{\omega }_{4}^{24}\simeq \widehat{\omega }
_{4}^{33}\simeq \widehat{\omega }_{4}^{41}\simeq \widehat{\omega }
_{4}^{44}\simeq O\left( \widetilde{\Theta }^{2}\right)
\end{eqnarray}

\begin{equation}
\widehat{\omega }_{1}^{14}=-\widehat{\omega }_{1}^{41}=- \frac{\beta 
\widehat{t}^{\beta -1}}{t_{0}}+O\left( \widetilde{\Theta }^{2}\right)
\label{B-29}
\end{equation}

\begin{equation}
\widehat{\omega }_{1}^{23}=-\widehat{\omega }_{1}^{32}=-\frac{i\widetilde{
\Theta }}{2}\beta ^{2}\frac{\widehat{r}}{t_{0}}h\widehat{t}^{2\beta -2}\sin
\theta +O\left( \widetilde{\Theta }^{3}\right)  \label{B-30}
\end{equation}

\begin{equation}
\widehat{\omega }_{1}^{31}=-\widehat{\omega }_{1}^{13}=-\frac{i\widetilde{%
\Theta }}{2}\beta ^{2}\frac{ \widehat{r}}{t_{0}}h\widehat{t}^{2\beta -2}\cos
\theta +O\left( \widetilde{ \Theta }^{3}\right)  \label{B-31}
\end{equation}

\begin{equation}
\widehat{\omega }_{2}^{12}=-\widehat{\omega }_{2}^{21}=- 1+O\left( 
\widetilde{\Theta }^{2}\right)  \label{B-32}
\end{equation}

\begin{equation}
\widehat{\omega }_{2}^{23}=-\widehat{\omega }_{2}^{32}=- \frac{i\widetilde{
\Theta }}{4}\left( -1+\widehat{t}^{2\beta -2}\beta ^{2}h^{2}\widehat{r}
^{2}\right) \cos \theta +O\left( \widetilde{\Theta }^{3}\right)  \label{B-33}
\end{equation}

\begin{equation}
\widehat{\omega }_{2}^{24}=-\widehat{\omega }_{2}^{42}=- \beta \widehat{r}h 
\widehat{t}^{\beta -1}+O\left( \widetilde{\Theta }^{2}\right)  \label{B-34}
\end{equation}

\begin{equation}
\widehat{\omega }_{2}^{13}=\frac{i\widetilde{\Theta }}{4}\left( 1+\widehat{t}
^{2\beta -2}\beta ^{2}h^{2}\widehat{r}^{2}\right) \sin \theta +O\left( 
\widetilde{\Theta }^{3}\right)  \label{B-35}
\end{equation}

\begin{equation}
\widehat{\omega }_{2}^{31}=\frac{i\widetilde{\Theta }}{4}\left( 1+2\widehat{%
t }^{2\beta -2}\beta ^{2}h^{2}\widehat{r}^{2}\right) \sin \theta +O\left( 
\widetilde{\Theta }^{3}\right)  \label{B-36}
\end{equation}

\begin{equation}
\widehat{\omega }_{2}^{34}=-\frac{i\widetilde{\Theta }}{4}\beta h\widehat{r} 
\widehat{t}^{\beta -1}\left( 1+2\widehat{t}^{2\beta -2}\beta ^{2}h^{2} 
\widehat{r}^{2}\right) +O\left( \widetilde{\Theta }^{3}\right)  \label{B-37}
\end{equation}

\begin{equation}
\widehat{\omega }_{2}^{43}=-\frac{i\widetilde{\Theta }}{4}\beta h\widehat{r} 
\widehat{t}^{\beta -1}\left( 1+\widehat{t}^{2\beta -1}\beta ^{2}h^{2} 
\widehat{r}^{2}\right) +O\left( \widetilde{\Theta }^{3}\right)  \label{B-38}
\end{equation}

\begin{equation}
\widehat{\omega }_{3}^{11}=\frac{i}{8}\widetilde{\Theta }\sin 2\theta
+O\left( \widetilde{\Theta }^{3}\right)  \label{B-39}
\end{equation}

\begin{equation}
\widehat{\omega }_{3}^{12}=\frac{i\widetilde{\Theta }}{4}\left( \cos
^{2}\theta -\widehat{t}^{2\beta -2}\beta ^{2}h^{2}\widehat{r}^{2}\sin
^{2}\theta \right) +O\left( \widetilde{\Theta }^{3}\right)  \label{B-40}
\end{equation}

\begin{equation}
\widehat{\omega }_{3}^{13}=-\widehat{\omega }_{3}^{31}=- \sin \theta
+O\left( \widetilde{\Theta }^{2}\right)  \label{B-41}
\end{equation}

\begin{equation}
\widehat{\omega }_{3}^{14}=\widehat{\omega }_{3}^{41}=-\frac{i}{8}\widetilde{
\Theta }\beta h\widehat{r}\widehat{t}^{\beta -1}\sin 2\theta +O\left( 
\widetilde{\Theta }^{3}\right)  \label{B-42}
\end{equation}

\begin{equation}
\widehat{\omega }_{3}^{21}=-\frac{i\widetilde{\Theta }}{4}\left( 1+2\widehat{
t}^{2\beta -2}\beta ^{2}h^{2}\widehat{r}^{2}\right) \sin ^{2}\theta +O\left( 
\widetilde{\Theta }^{3}\right)  \label{B-43}
\end{equation}

\begin{equation}
\widehat{\omega }_{3}^{22}=-\frac{i\widetilde{\Theta }}{8}\left( 1+3\widehat{
t}^{2\beta -2}\beta ^{2}h^{2}\widehat{r}^{2}\right) \sin 2\theta +O\left( 
\widetilde{\Theta }^{3}\right)  \label{B-44}
\end{equation}

\begin{equation}
\widehat{\omega }_{3}^{23}=-\widehat{\omega }_{3}^{32}=- \cos \theta
+O\left( \widetilde{\Theta }^{2}\right)  \label{B-45}
\end{equation}

\begin{equation}
\widehat{\omega }_{3}^{24}=\frac{i\widetilde{\Theta }}{4}\beta h\widehat{r} 
\widehat{t}^{\beta -1}(1+2\widehat{t}^{2\beta -2}\beta ^{2}h^{2}\widehat{r}
^{2})\sin ^{2}\theta +O\left( \widetilde{\Theta }^{3}\right)  \label{B-46}
\end{equation}

\begin{equation}
\widehat{\omega }_{3}^{33}=i\frac{\widetilde{\Theta }}{2}\beta ^{2}h^{2} 
\widehat{r}^{2}\widehat{t}^{2\beta -2}\sin 2\theta +O\left( \widetilde{
\Theta }^{3}\right)  \label{B-47}
\end{equation}

\begin{equation}
\widehat{\omega }_{3}^{34}=-\widehat{\omega }_{3}^{43}=- \beta h\widehat{r} 
\widehat{t}^{\beta -1}\sin \theta +O\left( \widetilde{\Theta }^{2}\right)
\label{B-48}
\end{equation}

\begin{equation}
\widehat{\omega }_{3}^{42}=-\frac{i\widetilde{\Theta }}{4}\beta h\widehat{r} 
\widehat{t}^{\beta -1}\left[ \cos ^{2}\theta -\widehat{t}^{2\beta -2}\beta
^{2}h^{2}\widehat{r}^{2}\sin ^{2}\theta \right] +O\left( \widetilde{\Theta }
^{3}\right)  \label{B-49}
\end{equation}

\begin{equation}
\widehat{\omega }_{3}^{44}=\frac{i\widetilde{\Theta }}{8}\beta ^{2}h^{2} 
\widehat{r}^{2}\widehat{t}^{2\beta -2}\sin 2\theta +O\left( \widetilde{
\Theta }^{3}\right)  \label{B-50}
\end{equation}

\begin{equation}
\widehat{\omega }_{4}^{23}=-\widehat{\omega }_{4}^{32}=- i\frac{\widetilde{
\Theta }}{2}\beta ^{2}\left( \beta -1\right) \frac{\widehat{r}^{2}}{t_{0}}
h^{2}\widehat{t}^{2\beta -3}\sin \theta +O\left( \widetilde{\Theta }
^{3}\right)  \label{B-51}
\end{equation}

\begin{equation}
\widehat{\omega }_{4}^{34}=-i\frac{\widetilde{\Theta }}{2}\beta \left( \beta
-1\right) \frac{\widehat{r}}{t_{0}}h\widehat{t}^{\beta -2}\cos \theta
+O\left( \widetilde{\Theta }^{3}\right)  \label{B-52}
\end{equation}

\begin{equation}
\widehat{\omega }_{4}^{43}=0.
\end{equation}

\end{document}